# INFLUENCE OF WETTABILITY ON LIQUID WATER TRANSPORT IN GAS DIFFUSION LAYER OF PROTON EXCHANGE MEMBRANE FUEL CELLS (PEMFC)

**H.CHRAIBI, L.CEBALLOS[*], M.PRAT, M.QUINTARD**
Institut de Mécanique des Fluides de Toulouse,UMR CNRS-INP/UPS No. 5502
Avenue du Professeur Camille Soula, 31400 Toulouse, France

**A.VABRE**
[*]CEA Saclay, Bâtiment 516, 91191 Gif-Sur-Yvette, France

**ABSTRACT**
Water management is a key factor that limits PEFC's performance. We show how insights into this problem can be gained from pore-scale simulations of water invasion in a model fibrous medium. We explore the influence of contact angle on the water invasion pattern and water saturation at breakthrough and show that a dramatic change in the invasion pattern, from fractal to compact, occurs as the system changes from hydrophobic to hydrophilic. Then, we explore the case of a system of mixed wettability, i.e. containing both hydrophilic and hydrophobic pores. The saturation at breakthrough is studied as a function of the fraction of hydrophilic pores. The results are discussed in relation with the water management problem, the optimal design of a GDL and the fuel cell performance degradation mechanisms. We outline how the study could be extended to 3D systems, notably from binarised images of GDLs obtained by X ray microtomography.

## 1. INTRODUCTION

Water management is a key factor that limits PEFC's performance and it is now well admitted that a detailed understanding of liquid water transport in the gas diffusion layer (GDL) is necessary to facilitate development of techniques and GDL materials and to mitigate flooding and associated mass transport losses, e.g. [1]. Also, a better understanding of the effect of wettability and its possible evolution during PEMFC operation can help understanding the degradation processes that reduce PEFC's performance.

From a modelling standpoint, it is however difficult to gain insights into the effect of wettability from the traditional macroscopic approach based on two-phase Darcy's law. Macroscopic models cannot incorporate the GDL morphology and require material-specific capillary pressure –liquid saturation and relative – permeability –liquid saturation relationships that are not easy to determine for a thin medium such as a GDL. Furthermore, the quality of approximation proposed by the macroscopic model remains to be investigated owing to the lack of length scale separation, at least along the depth of GDL, i.e. a GDL is only a few pores deep whereas the continuum approach to porous media requires a priori that the pore size be very small compared to the porous domain size.

As pointed out in [1] or [2] among others, pore-scale studies are much more adapted for studying the impact of wettability and microstructure. Although direct simulations of two phase flows at the pore-scale are possible with lattice-Boltzmann methods or by solving the Navier-Stokes equations using more classical discretisation techniques, their interest is severely limited by their computational cost, which for example make impossible a statistical study. An attractive possibility in this context is represented by the so-called pore network models (PNM), e.g. [2] and references therein, which allow considering a sufficiently large number of pores into the computational domain. PN Models can be used for different objectives: as a tool for understanding the physics of two-phase flow (including possibly evaporation) or as a tool for predicting the constitutive relations evoked before (capillary pressure, relative permeabilities) or other macroscale transport properties (permeability, effective diffusion coefficient, etc).

Our objective in this context is to study the impact of wettability on water invasion in a fibrous medium using a PNM. To this end, we consider a model fibrous medium made of disks of random diameters placed on a square lattice. Noting that water invasion in GDL is generally dominated by capillary effects, we concentrate on the quasi-static limit and model the displacement of the water-air interface for any contact angle. We begin with the case of spatially uniform wettability and study the evolution of the invasion pattern as a function of contact angle.

Then we study the effect of mixed wettability. We start with a medium of uniform wettability (contact angle 110°) and study the evolution of saturation at breakthrough as a function of the fraction of hydrophilic pores (contact angle of 80°, which is seen here as the contact angle on carbon). In this study, the hydrophilic pores are distributed randomly and form disconnected clusters of hydrophilic pores until a hydrophilic pore percolation threshold is reached. The impact of the results on the fuel cell operation and the water management problem is discussed.

Finally, we outline how this type of model could be extended to 3D simulations and combine with X-ray microtomographies of real GDLs so as to approach as much as possible representative GDL microstructures through the generation of topologically equivalent pore-network structure.

## 2. MODEL FIBROUS MEDIUM, INVASION ALGORITHM

As depicted in Figure 1, the model used in this study is a 2D array of disks (representing the fibers) with random radii. A distinguishing advantage of this model is to allow for a full solution of the interface shape. In particular, it is possible to study in detail the influence of contact angle on the invasion pattern. In this model, the liquid-gas interface consists of circular arcs connecting disks. The disks are placed on a square lattice with lattice constant $a$. There are $L$ disks per row and $L$ rows on the lattice. Disk radii are distributed randomly according to an uniform distribution law in the range $[r_{min}, r_{max}]$. Initially, the pores of the model porous medium are saturated with gas.

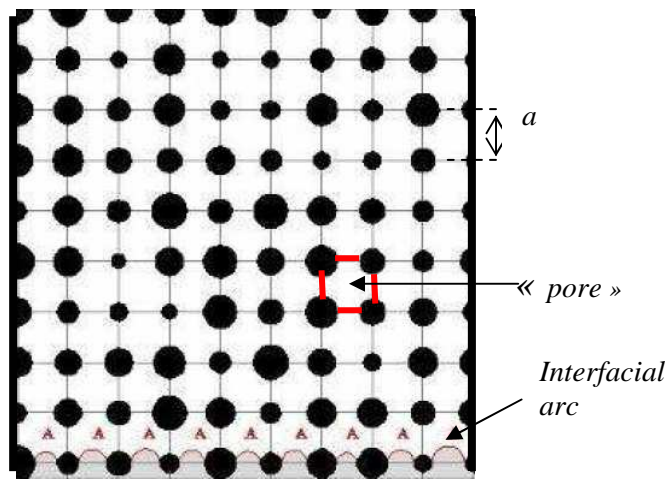

Figure 1 Model fibrous medium formed by a regular array of disks of random diameter. Distribution of fluids at $t = 0$ (invading fluid in grey, displaced fluid in white). The interface is visible as a series of arcs joining the bottom row of disks. Water is very slowly injected from the bottom edge of system and displaces the air in place, which escapes from the top edge.

Water is injected from the bottom edge and the simulation of the invasion is performed until liquid water breaks out from the top edge of the pore network (breakthrough). As discussed in [2], the capillary number characterizing the displacement, i.e. the invasion of the GDL by liquid water, is expected to be very small under standard fuel cell operating conditions. As a result, the displacement can be considered as quasi-static, i.e. simulated as a sequence of successive capillary equilibriums.

The wettability of the system is characterized by the contact angle $\theta$ measured in the dense phase, i.e. liquid water in the present context. This contact angle can be interpreted as the advancing contact angle since water is the invading phase.

At each step of the invasion, one "interfacial pore" of the system is invaded. A "pore" is defined as the region of the pore space located inside the square formed by the segments joining two neighbour disks on the square lattice (see Fig. 1). An interfacial pore is a pore not yet invaded containing at least one arc. Interfacial pores are indicated with a "A" in Fig.1. To determine which interfacial pore should be invaded at a given step of the invasion, we determine the "critical" invasion curvature radius of each interfacial arc. As discussed in more details in [2], this consists in studying the arc curvature radius as the arc moves between the two disks it intercepts. As discussed in [2], there are three main local growth mechanisms controlling the invasion. The first is called "burst". It corresponds to the burst of the meniscus when the local capillary pressure exceeds the maximum local capillary pressure compatible with a stable arc joining the pair of disks under consideration, see [3] or [4] for more details. We note $r_b$ the arc curvature radius corresponding to the last arc stable before burst. The local capillary pressure corresponding to the arc burst may be not reached because another event occurs before: the considered arc touches a third disk during its growth, or coalesces with another arc. If this happens, we compute the arc radius of curvature corresponding to each of these local events with $r_t$ the radius corresponding to the touch event and $r_c$ the radius corresponding to the coalescence event, if any. These radii are computed as functions of contact angle, disks radii and distances between disks, [3], [4]. We affect to each interfacial arc the radius $r_a = \max(r_b, r_t, r_c)$. This gives a hierarchy of radii. At each step of the invasion, the interfacial pore associated with the arc(s) corresponding to max $(r_a)$, i.e. the minimum invasion threshold local capillary pressure, is invaded. The corresponding "unstable" arc(s) is (are) suppressed from the list of interfacial arcs and new arcs are created at the periphery of the invaded pore. This procedure is repeated until breakthrough. Again the interested reader can refer to [2] or [4] for more details on the invasion procedure.

## 3. INFLUENCE OF CONTACT ANGLE ON INVASION PATTERN

Figure 2 shows the evolution of the invasion pattern as a function of the contact angle $\theta$. Liquid water being the invading fluid, a hydrophilic fibrous medium corresponds to $0 \leq \theta < 90°$ whereas a hydrophobic one corresponds to $90° < \theta \leq 180°$. As can be seen from Fig.2, the wettability has a major impact on the invasion pattern. For large contact angles ($\Leftrightarrow$ very hydrophobic GDL), capillary fingering is obtained. For contact angles lower than about $100°$ ($\Leftrightarrow$ hydrophilic GDL), a compact pattern characterized by an almost flat invasion front is obtained. It is interesting to note that the transition between the two patterns is sharp but does not occur exactly and abruptly at 90°. The transition occurs in fact over a narrow range of contact angles, $[80°, 100°]$ in this example.

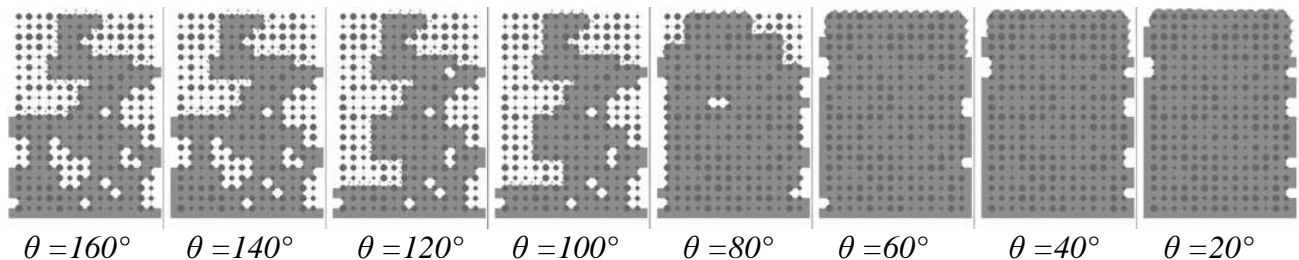

$\theta = 160°$    $\theta = 140°$    $\theta = 120°$    $\theta = 100°$    $\theta = 80°$    $\theta = 60°$    $\theta = 40°$    $\theta = 20°$

Figure 2. Evolution of invasion pattern at breakthrough in a model fibrous medium as a function of contact angle $\theta$. Invading phase (liquid water) in grey, displaced phase (air) in white

As discussed in more details in [2], the change in the invasion pattern can be explained by the change in the interface local growth mechanisms. Burst is the dominant mechanism for high contact angle, and the invasion can be simulated using the classical invasion percolation (IP) algorithm, [5]. For contact angles lower than about $100°$, coalescence of menisci is the dominant mechanism and the invasion can be simulated very simply by a flat travelling front (note however that for sufficiently low contact angles below $100°$, film effects can complicate the invasion with certain pore geometries, e.g. [6] and references therein).

The patterns shown in Fig.2 indicate that rendering the GDL hydrophobic is a good option. Large regions of the pore space remains accessible to the gas phase up to the CL/GDL interface in the capillary fingering regime that characterizes water invasion in a hydrophobic GDL. In contrast, the compact invasion in a hydrophilic GDL is very detrimental to the gas access to the GDL/CA interface. As illustrated in Fig.3, the liquid saturation $S$ at breakthrough (when the liquid reaches the channel) in the hydrophobic GDL is much lower than for a hydrophilic one (where $S \sim 1$). The saturation $S$ represents the fraction of the pore volume occupied by liquid water at breakthrough. When $S \sim 1$, all pores are full of water at breakthrough, which represents the worse case since there is no pore pathway available for the gas across the GDL.

Thus our results provide a pore–scale physical explanation to the fact that it is advantageous to use hydrophobic GDLs.

As mentioned in the introduction, most models of two –phase flows in GDLs are based on the traditional two-phase generalized Darcy's law and the concept of macroscopic capillary pressure. Contrary to what can be found in many papers on the modelling of two-phase flow in GDL(s), the fact that the invasion regime (compact vs capillary fingering) is completely different depending on the wettability of the GDL indicates that it is not correct to use basically the same mathematical expressions for the Leverett function (i.e. the capillary pressure-liquid saturation relationship) and the relative permeabilities whatever the wettability of the GDL, i.e. just changing the value of the contact angle in these expressions is not correct.

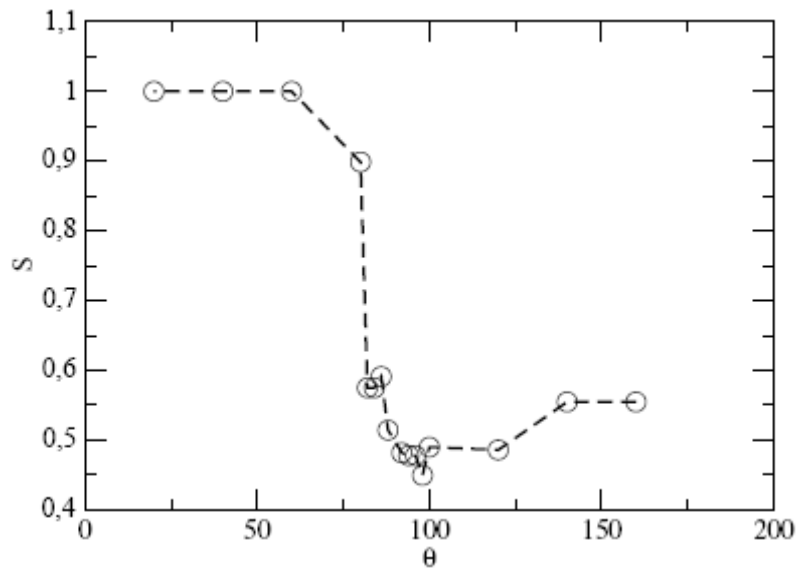

Figure 3. Evolution of overall water saturation at breakthrough, corresponding to patterns shown in Fig.2, as a function of contact angle $\theta$.

## 3. SYSTEM OF MIXED WETTABILITY

The results presented in section 2 were for a system of spatially uniform contact angle. However, the load in PTFE can vary, typically from 5 to 30 wt%, and it is likely that the distribution of PTFE is not uniform within a GDL. Although the distribution of PTFE on the fibers forming the GDL is still unclear, a simple model is to assume that some pores of the GDL are hydrophilic (i.e. $\theta \approx 80°$ on a carbon fiber) whereas the others are hydrophobic (i.e. $\theta \approx 110°$ on a perfectly teflonized fiber), e.g. [7]. In what follows, the fraction of hydrophilic pores ($\theta \approx 80°$) is denoted by $p$. Hence, $p$ = number of hydrophilic pores / total number of pores, 1-$p$ is the fraction of hydrophobic pores ($\theta \approx 110°$). Extending the invasion model presented in section 2 to the case where the contact angle is 80° for a fraction of pores and 110° for the other pores is straightforward. The curvature radii mentioned in section 2 are simply computed as a function of the local contact angle associated with the considered interfacial pore. In order to get some insights into the average behaviour of the system, we have in principle to take into account two sources of randomness. The first one is the randomness associated with the disk radii. The second one is the randomness introduced by the random distribution of the hydrophilic pores (the hydrophilic pores are selected at random, the remaining pores are specified as hydrophobic). Thus, computing the average behaviour implies to combine two ensemble averages. This can be achieved by generating many realizations of the disk radii distribution and then, for each of this realization, by generating many realizations of the pore wettability distributions. For simplicity, however, the ensemble averaging was carried out only as regards the randomness of pore wettability. Hence we look at the impact of $p$ for a given realization of disk array (note that only one realization was also considered in section 2). Of particular interest as regards the water management problem is the liquid water saturation $S$ at breakthrough. The evolution of $S$ as a function of $p$ is shown in Fig.4a. These results have been obtained for a 40x40 disks array. For each value of $p$ considered, $S$ is the average over 100 realizations of the pore wettability distributions.

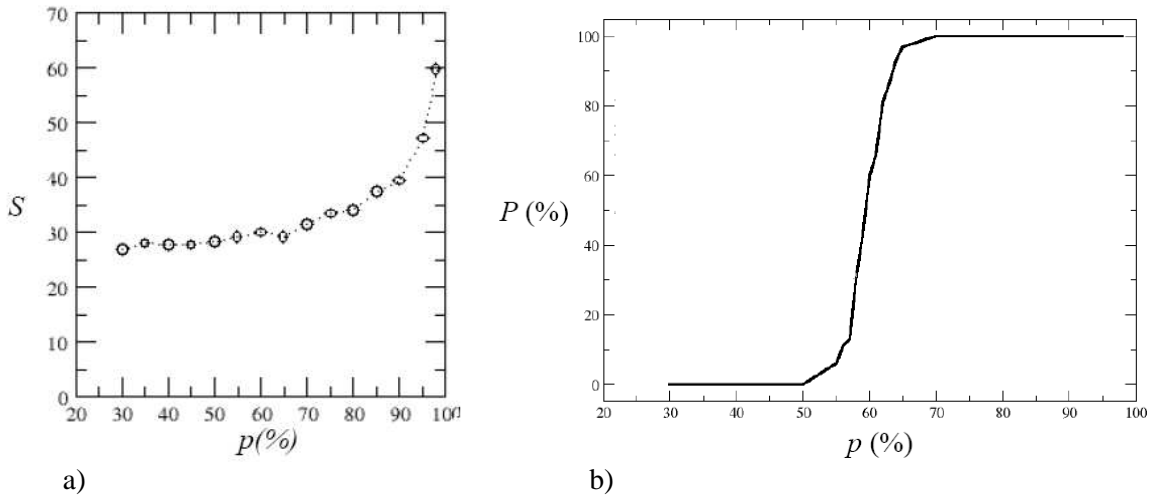

Figure 4. a) Evolution of liquid saturation at breakthrough as a function of hydrophilic pore fraction
b) percolation probability of hydrophilic pore subset as a function of hydrophilic pore fraction.

As in any classical percolation problem, [8], the hydrophilic pores form isolated pore clusters for sufficiently low values of $p$. When $p$ is sufficiently large, there exists a cluster of hydrophilic pores spanning the system. The critical pore fraction below which such a cluster does not exist is the percolation threshold $p_c$ ($p_c \approx 0.59$ for a very large square lattice). In a finite size system, the percolation transition is not sharp and some realisations are percolating below $p_c$ whereas others are percolating for values of $p$ greater than $p_c$. This is illustrated in Fig.4b, which shows the evolution of the percolation probability $P$ as a function of $p$ (for a given value of $p$, $P$ is the fraction of percolating realizations among the 100 generated realizations). As can be seen from Fig.4b, our results are fully consistent with the value 0.59 for $p_c$.

Since liquid water favours hydrophilic pores, liquid water invasion takes place through a path of hydrophilic pores above $p_c$ and the hydrophobic pores are left free of water, i.e. available for gas transport. Below $p_c$, some hydrophobic pores are necessarily invaded since the hydrophilic pores do not form a percolating cluster. However, a hydrophilic pore cluster is systematically fully invaded when liquid water meets such a hydrophilic cluster during the invasion. We know from section 2 that the invasion pattern and the local invasion mechanisms are significantly different for $\theta \approx 80°$ and $\theta \approx 110°$. Interestingly, this does not have, however, a great impact on the saturation at breakthrough over a large range of $p$, as show in Fig.4a. This can be qualitatively understood as follows. For $p$ close to zero, the water invasion process is essentially an invasion percolation (IP) process as discussed in section 2. As it is well known, the subset of pores invaded at breakthrough as a result of the IP process is very similar to a ordinary percolation cluster, i.e. the percolating cluster formed by the hydrophilic pores at $p_c$. Hence, since the pore size distribution is generally narrow in a GDL, we expect that $S(0) \approx S(p_c)$. For intermediate $p$ between 0 and $p_c$, the invaded pore subset can be envisioned as a chain of invaded hydrophilic pore clusters connected by paths of hydrophobic pores, i.e. a structure close to a percolation cluster at $p_c$.

The main practical conclusion from the results shown in Fig.4, is that the PTFE load has little impact on the water management problem as long as the fraction of hydrophilic pores is not significantly larger than the percolation threshold of the system.

As also pointed out in [7] from a limited series of 3D simulations, our results suggest the possibility of optimizing the GDL transport properties by controlling the wettability of fibres forming the GDL. In our example, the water invasion pathway through a percolating cluster of hydrophilic pores above $p_c$ was obtained by assuming a random distribution of hydrophilic pores. In a 3D system, this leads to separate transport paths for the liquid and the gas across the GDL if $p_c \leq p < 1 - p_c$ with liquid flowing through a cluster of hydrophilic pores and the gas through the remaining pores. If can be anticipated that distributions of hydrophilic pores other than purely random would lead to a more efficient water management.

## 4. DISCUSSION

As discussed above, elucidation of pore-level physics of liquid water transport in GDL is essential not only to understand the effect of wettability on flooding phenomena but also to design optimized GDLs through for example the consideration of GDLs of controlled mixed wettability.

Also, the type of simulations presented in this paper can be helpful for studying the mechanisms of degradation of fuel cell performances. A possible mechanism is the evolution of wettability properties of GDLs.

A possible scenario is the increase of the fraction of hydrophilic pores over time. The results shown in Figure 4 suggest that the change in this fraction of pores should be very significant for affecting the saturation at breakthrough, i.e. a very significant change in the fraction of hydrophilic pores is needed for affecting the fuel cell performances in this respect (assuming that the initial fraction of hydrophilic pores is not too close to the percolation threshold). This was obtained by assuming a significant local change in the contact angle (from 110° to 80 °) in some of the pores.

Another scenario is a more gradual change in the contact angle but in every hydrophobic pore. In this case, the results presented in section 2, see Figure 3, suggest a more dramatic conclusion. As shown in section 2, the transition from a fractal pattern to a compact pattern as the contact angle is varied is rather sharp. In this example, the sharp transition occurs for contact angles close to 90°, i.e. relatively close to the value of 110° generally considered for water on a teflonized surface. Hence, this suggests that a relatively small change in the contact angle of "hydrophobic" pores may greatly affect the invasion pattern and therefore the access of reactants to the active layer since the fuel cell operates close to the pattern transition. This clearly indicates that the impact of a change in wettability can be subtle in systems of intermediate wettability like a GDL. To discriminate between various possible scenarios, experimental investigations are certainly needed to characterize the fibre wettability inside a GDL and, possibly, the change in wettability over time.

We have only considered a somewhat idealized 2D model of fibrous medium. Although many insights can be gained from this simple model, the consideration of more realistic model of GDLs is needed. In particular, it is well known that the percolation properties of a 3D system are significantly different from a 2D case. Also, we have only discussed the saturation at breakthrough. The computation of transport properties from realistic microstructure (gas effective diffusion coefficient, relative permeabilities, etc) is also needed for optimizing the design of a GDL. This makes sense only in 3D.

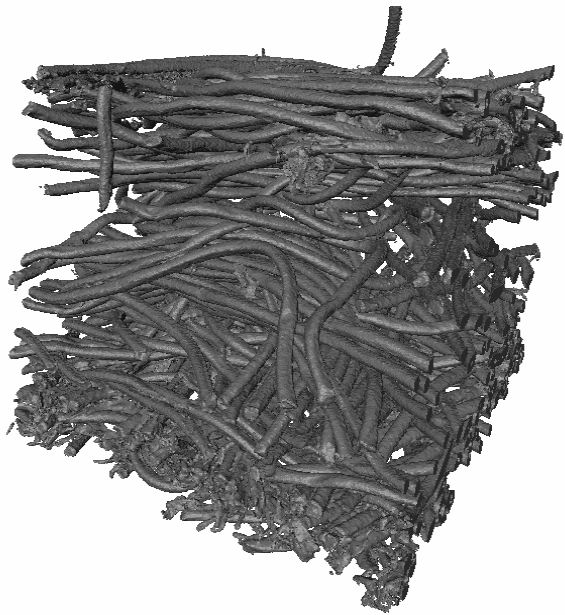

Figure 5 Example of image of GDL microstructure obtained by X ray microtomography

Although consideration of simple 3D pore network models is certainly helpful, e.g. [7], it is naturally desirable to approach real GDL microstructures as much as possible. To this end, we are currently developing X-ray microtomographies of real GDLs. Figure 5 shows a representative image of GDL obtained with this technique. The size of the sample shown in Figure 5 is 300x300x300µm whereas the size of the voxel is 0.45x0.45x0.45 µm. It should be noted that this type of image is not easy to obtain owing to the relatively weak X ray absorption by GDL fibers and the high spatial resolution that is needed. The details on these aspects will be presented elsewhere. In parallel, works is in progress for extracting a pore-network structure from this type of image. Naturally, this pore network structure will be useful not only for the simulation of two-phase flows, in the spirit of what has been shown in this paper, but also for characterizing the GDL macroscopic transport properties (permeability, effective diffusion tensors, relative permeabilities,..). Works in this direction are in progress.

**ACKNOWLEDGEMENT:**
Financial support from GIP ANR (project ANR-06-PANH-022-02 "Chameau") is gratefully acknowledged